# Inhomogeneous Charge State in HTSC Cuprates and CMR Manganites

T. Egami

*Department of Materials Science and Engineering and Laboratory for Research on the Structure of Matter, University of Pennsylvania, Philadelphia, PA 19104*

**Abstract**: Recent measurements of neutron elastic and inelastic scattering suggest that the charge states are spatially inhomogeneous at two lengthscales, atomic and nanometer scales, in both the high-temperature superconducting (HTSC) cuprates and colossal magnetoresistive (CMR) manganites. We suggest that the two-phonon mechanism that controls the charge localization in CMR manganites is also at work in HTSC cuprates, and may hold a key to understanding the mechanism of superconductivity.

Key words: neutron scattering, charge inhomogeneity, phonons, CMR manganites

## 1. Introduction

Generally local lattice distortion and charge inhomogeneity are considered to be harmful to superconductivity, including high-temperature superconductivity (HTSC). Thus in spite of numerous reports of local lattice distortions in cuprates [1] the majority view has been that they are unimportant to the mechanism of superconductivity, and are merely the consequences of strong interactions and short coherence length. However, the perception appears to be changing due to the observation of the spin-charge stripes in non-superconducting cuprates with 1/8 charge density and the conjecture that dynamic stripes might be present in the superconducting state [2]. In this paper we discuss recent results by neutron inelastic scattering that indicate the presence of charge inhomogemeities at two lengthscales, atomic and nanometer scales, which are in many ways similar to those in the manganites that show colossal magnetoresistance (CMR). We suggest that the two-phonon mechanism of charge localization in CMR manganites must be in operation also in cuprates, and such phonon involvement may be an integral part of the HTSC phenomenon.

## 2. Inelastic Neutron Scattering

We have made a series of inelastic neutron scattering measurements on the high-energy LO phonons in cuprates, $La_{1.85}Sr_{0.15}CuO_4$ [3] and $YBa_2Cu_3O_{6+x}$ [4,5]. The pho-non branch we studied was the Cu-O bond-stretching mode as shown in Fig. 1. It mainly involves oxygen motion, and at the zone-center, $q = 0$, it is the ferroelectric mode, while at the zone-edge, $q = (\pi, 0, 0)$ in the unit of $1/a$ ($a$ is the Cu-Cu distance), it is a half-breathing mode. This mode induces the hybridization between the Cu-$d$ and O-$p$ orbitals to change and results in charge transfer. For this reason it is expected to interact strongly with charge carriers [6,7]. Indeed this is the only mode that shows strong softening near the zone-edge upon doping [8], and the kink in the electron dispersion determined by photoemission is likely due to this phonon mode [9]. The neutron results revealed a rather anomalous nature of this branch:

1. At low temperatures the branch splits into two, a high-energy branch and a low energy-branch, both of which show little dispersion.
2. The splitting of the phonon dispersion becomes weak at higher temperatures. In YBa$_2$Cu$_3$O$_{6.95}$ the temperature dependence of the change in the dispersion agrees with the superconducting order parameter.
3. As the doped charge density is increased in YBCO, one would expect the LO phonon dispersion to soften gradually near the zone-edge. Surprisingly the energy at the zone-edge does not change much with doping and the phonon dispersion is always split into two. But the intensity is transferred from the high energy-branch to the low-energy branch with increased doping.

The point 1 is best explained by doubling of the unit cell that halves the Brillouin zone, such as the charge pattern shown in Fig. 2 that has the periodicity of $2a$ [3]. The usual stripe pattern with the periodicity of $4a$ fails to account for the dispersion. The $2a$ periodicity appears to reflect the underlying Peierls instability, and may give rise to the coupled ladder state [10]. As point 2 shows this charge pattern is temperature dependent. The point 3 is best interpreted in terms of an inhomogeneous structure; microscopic segregation into charge-rich and charge-poor regions. The high-energy branch must be associated with charge-poor regions, since it agrees with the dispersion in the undoped sample [8], and the low-energy branch with charge-rich regions, since its intensity grows with doping. The lengthscale for the segregation must be a nanometer at most; otherwise the $q$ dependence of the phonon structure factors for the two branches must be similar. This micro-phase segregation was seen also by a neutron scattering PDF study [11]. Thus the results suggest that the charge state of cuprates is inhomogeneous at two lengthscales, atomic and nanometer scales.

### 3. Polaron stability in CMR manganites

In manganites charges can become localized forming polarons, depending on temperature and composition [12,13]. Nanometer-scale charge inhomogeneity is also seen, even in the metallic state just above the insulator-to-metal transition, suggesting the charge segregation tendency [14,15]. There is a striking parallelism between polaron formation in manganites (0-d) and stripe formation in cuprates (1-d) as the mechanism of charge localization, and the presence of nanometer charge inhomogeneity in both. It is well understood that, in manganites, charge localization is controlled by the balance between the localization forces (electron-lattice coupling and antiferromagnetic or orbital ordering) and the delocalizaton forces (kinetic energy and elastic energy) [12,13]. These forces can be affected by the ionic size of the A-site ions in the perovskite, $<r_A>$ [16].

While the standard explanation is that the ionic size modulates the bandwidth, this has been questioned by a number of results [15]. Our alternate interpretation is based upon the local structure of polarons [15,17]. In the case of simple lattice polarons the spatial extensions of the charge and the lattice distortion are about the same. However, in manganites and cuprates spins play a large role in confining the charge, so that a hole can be localized on a single Mn site, while the elastic field that a polaron generates is long-range. This long-range stress field increases the polaron self-energy by a factor of about 2, according to the continuum elasticity theory [18,19]. But if the value of $<r_A>$ is small enough the Mn-O-Mn bonds are buckled. Then, when the local Jahn-Teller distortion is removed within the polaron and the long Mn-O bond locally becomes short, this change can be accommodated by unbuckling the bond (Fig. 3) [17]. In other words this unbuckling screens the local polaronic strain, and the long-range field will not be generated. Thus the renormalization of the polaron energy due to the long-range field does not happen, reducing the polaron energy by a factor of 2 compared to the previous case. In manganites this change in the polaron self-energy apparently is sufficient to stabilize polaron when $<r_A>$ is small, and destabilize when $<r_A>$ is large. At the crossover polarons are marginally stable, giving rise to the CMR phenomenon.

## 4. Electron-phonon coupling in cuprates

In our view an exact parallel should be observed for cuprates [20]. An increase in the local hole density will cause the Cu-O bond to be reduced in length, or in the phonon language, excite the in-plane LO phonons. If the Cu-O-Cu bond is buckled, the in-plane LO phonons couple to the c-axis oxygen mode, as in Fig. 3. The coupling is proportional to the deviation of the Cu-O-Cu bond angle from $\pi$. This coupling will soften the in-plane mode, by the screening mechanism described above. There are two possibilities of the in-plane modes that are softened by this coupling. The first obvious choice is the stripe mode ($q = \pi/2$), since when the buckling is strong charges will be localized into the stripe state, driving the system insulating. This is consistent with the effect of the $CuO_6$ tilting [21], and the need of Nd in $(La,Nd)_{1.875}Sr_{0.125}CuO_4$ to produce the stripe structure [2], since $Nd^{3+}$ is smaller than $La^{3+}$ (1.27 vs. 1.36 Å [22]). The second possibility is the Peierls mode ($q = \pi$). Our neutron results strongly suggest that this mode is present, perhaps competing against the stripe mode and consequently not developing into the stable state. This could produce a local spin-singlet state, suppress the antiferromagnetic order, and promote superconductivity [10]. At the same time the two-phonon mixing could produce multi-band HTSC [23,24]. While this scenario is highly speculative at this moment, it explains quite well the anomaly in the pulsed neutron PDF at $T_C$ involving the out-of-plane oxygen motion [25], and the dependence of $T_C$ on the "micro-strain" proposed recently by Bianconi *et al*. [26]. Indeed a remarkable similarity between the phase diagram in Fig. 6 of Ref. 26 and that for manganites (Fig. 4) [15,17] suggests that similar mechanisms must be at work. In Ref. 26, however, the stripe mode itself is considered to generate HTSC. In the present scenario the stripe mode merely terminates HTSC by CDW formation, while the competition between the Peierls state and the stripe state plays a crucial role in HTSC. Further experimental works are being conducted to examine these scenarios.

## 5. Conclusions

In CMR manganites the stability of polarons is determined by the local structure, i.e. screening of the local in-plane LO phonons by the c-axis oxygen phonons. An exact parallel should hold in cuprates where the coupling between LO phonons and c-axis phonons could play a crucial role in determining the stability of the stripe structure, and possibly superconductivity.

**Acknowledgments:** The author is grateful to the collaborators of the phonon and CMR projects, R. J. McQueeney, M. Yethiraj, D. Louca, Y. Petrov, M. Arai, J. F. Mitchell, H. A. Mook, G. Shirane and Y. Endoh. He is also thankful to A. Bianconi, S. J. L. Billinge, A. R. Bishop, A. Bussmann-Holder, J. B. Goodenough, L. P. Gor'kov, V. Kresin, A. Lanzara, K. A. Müller, J. C. Phillips, S. Sachdev, Z.-X. Shen, S. R. Shenoy, and M. Tachiki for useful discussions. Research at the University of Pennsylvania was supported by the National Science Foundation through DMR96-28136.

Figure captions:

Figure 1. (a) High-energy LO phonons at the zone-center, and (b) at the zone-edge, $(\pi, 0, 0)$.

Figure 2. A possible charge pattern with the $2a$ periodicity [3]. Large circles denote oxygen, with different charge densities indicated by grades of lightness. For clarity we show the pattern with a long-range order, but in reality the pattern should be dynamic and short-range, with the correlation length about $20 \times 8$ Å.

Figure 3. Accommodation of the reduction in the local Mn-O distance due to the presence of a hole on Mn, (a) when the Mn-O-Mn bond is straight this causes the neighboring Mn-O bonds to stretch, (b) when the Mn-O-Mn bond is buckled, unbuckling will locally accommodate the change in the Mn-O distance, without producing a long-range stress field [13,15].

Figure 4. Phase diagram for the CMR phenomenon in $A_{1-x}A'_xMnO_3$, plotted for the ionic radius (9-coordinated), $<r_A>$, against the charge density, x [13,15]. When $<r_A>$ is greater than 1.24 Å polarons are not stable, resulting in the metallic state. When $<r_A>$ is less than 1.18 Å polarons are stable, making the system insulating. The CMR phenomenon is observed in the crossover region.

(a)

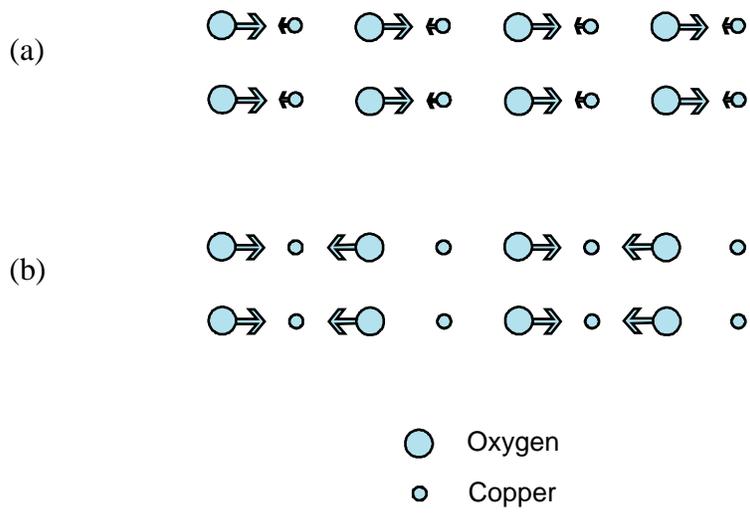

(b)

◯ Oxygen
∘ Copper

Fig. 1

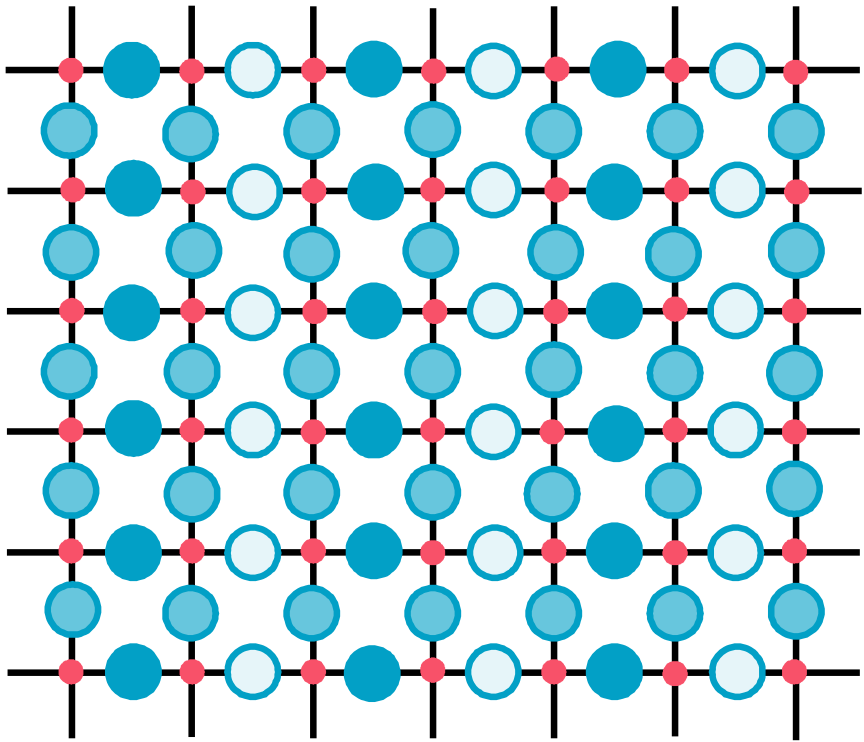

Fig. 2

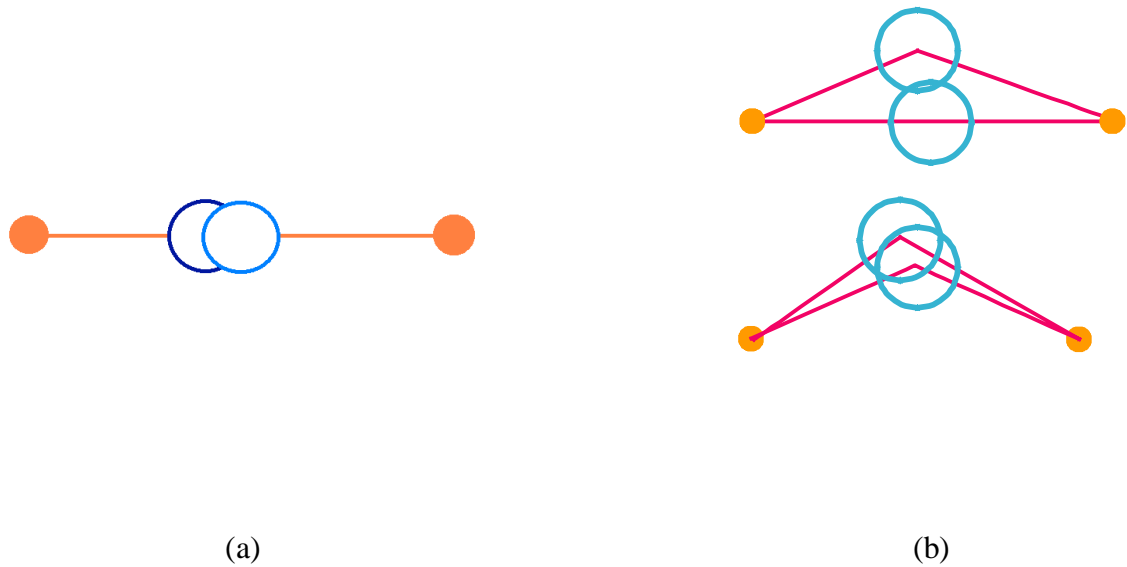

(a) (b)

Fig. 3

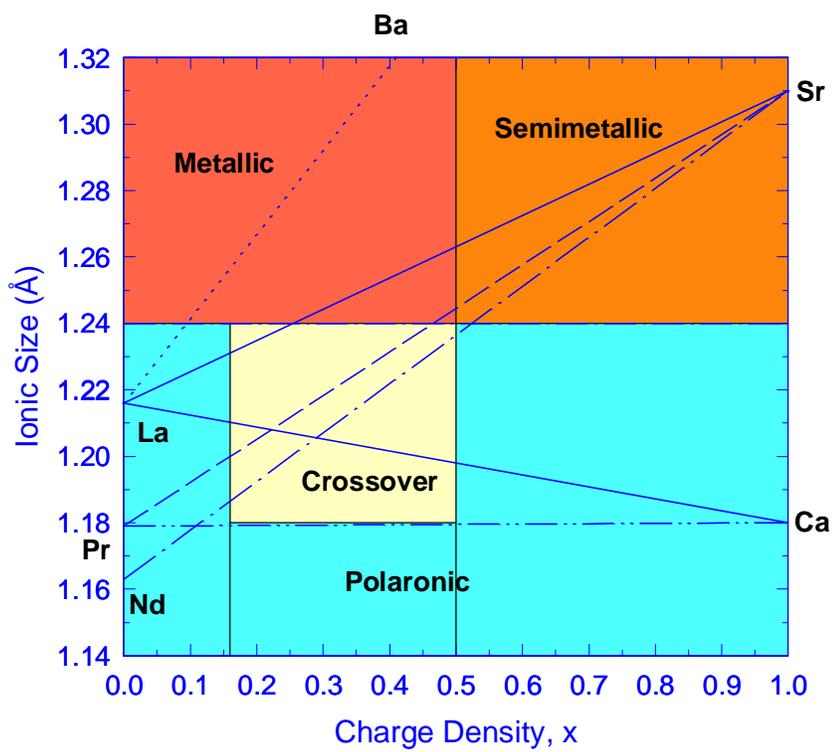

Fig. 4